\documentclass{arxiv}
\usepackage{eg2021}
 
\ConferencePaper        
\usepackage[T1]{fontenc}
\usepackage{dfadobe}  

\usepackage{cite}  
\electronicVersion
\PrintedOrElectronic
\ifpdf \usepackage[pdftex]{graphicx} \pdfcompresslevel=9
\else \usepackage[dvips]{graphicx} \fi

\usepackage{egweblnk} 

\usepackage{array, multirow, tikz, pgfplots, amsmath, amssymb, booktabs}
\pgfplotsset{compat=1.15} 
\usepackage{arydshln}
\usepackage{subcaption}

\newcommand\revised[1]{#1}

\title[Enabling Viewpoint Learning through Dynamic Label Generation]%
      {Enabling Viewpoint Learning through Dynamic Label Generation}

\author[M. Schelling et al.]
{\parbox
    {\textwidth}{\centering 
        M. Schelling$^1$\orcid{0000-0001-5294-4474}, 
        P. Hermosilla$^{1}$\orcid{0000-0003-3586-4741}, 
        P.-P. V\'azquez$^{2}$ \orcid{0000-0003-4638-4065}
        and T. Ropinski$^{1}$\orcid{0000-0002-7857-5512}
        }
        \\
{\parbox
    {\textwidth}{\centering 
        $^1$Ulm University, Germany\\
        $^2$Universitat Polit\`ecnica de Catalunya, Spain
       }
}
}

\begin{document}

\teaser{
 \includegraphics[width=\linewidth]{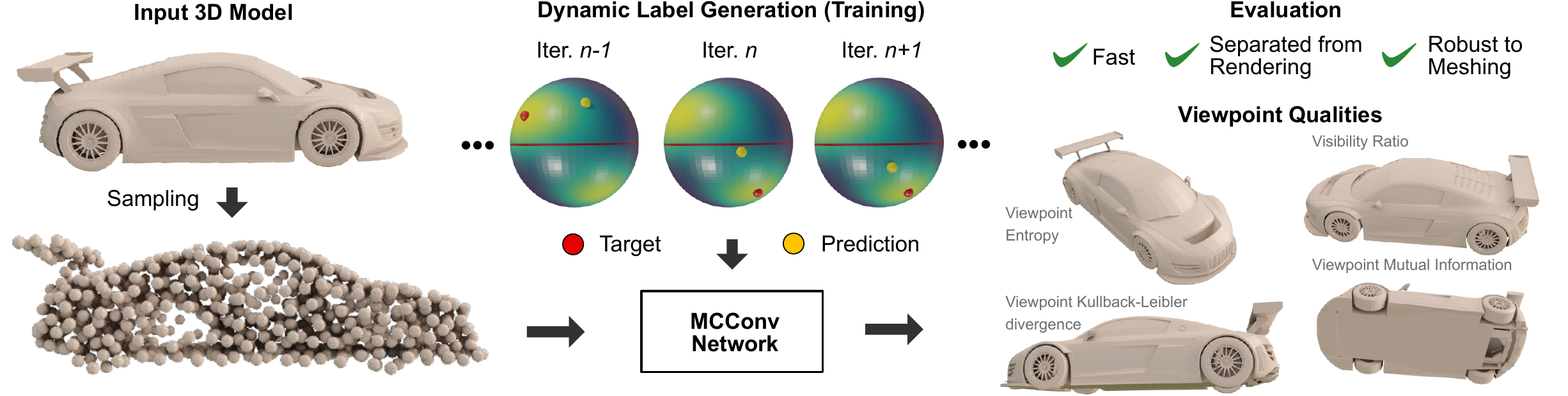}
 \centering
  \caption{We propose a new learning-based algorithm which is able to predict high quality viewpoints directly on 3D models. 
        The key to learning viewpoints is a novel approach to resolve label ambiguities, in the form of dynamic label generation, which adapts the network target during training,
        and enables our network to learn viewpoints for various viewpoint quality measures. 
        By learning solely on unstructured 3D point information, our approach is robust under mesh quality changes, and the viewpoint prediction is separated from the rendering process during evaluation.}
\label{fig:teaser}
}

\maketitle
\begin{abstract}
    Optimal viewpoint prediction is an essential task in many computer graphics applications. 
    Unfortunately, common viewpoint qualities suffer from two major drawbacks: dependency on clean surface meshes, which are not always available, and the lack of closed-form expressions, which requires a costly search involving rendering. 
    To overcome these limitations we propose to separate viewpoint selection from rendering through an end-to-end learning approach, whereby we reduce the influence of the mesh quality by predicting viewpoints from unstructured point clouds instead of polygonal meshes. While this makes our approach insensitive to the mesh discretization during evaluation, it only becomes possible when resolving label ambiguities that arise in this context. Therefore, we additionally propose to incorporate the label generation into the training procedure, making the label decision adaptive to the current network predictions. 
    We show how our proposed approach allows for learning viewpoint predictions for models from different object categories and for different viewpoint qualities. 
    Additionally, we show that prediction times are reduced from several minutes to a fraction of a second, as compared to state-of-the-art (SOTA) viewpoint quality evaluation. 
    We will further release the code and training data, which will to our knowledge be the biggest viewpoint quality dataset available.
\end{abstract}

\newcommand{\VR}{$\mathrm{VR}$}
\newcommand{\VE}{$\mathrm{VE}$}
\newcommand{\VKL}{$\mathrm{VKL}$}
\newcommand{\VMI}{$\mathrm{VMI}$}


\section{Introduction}\label{sec:intro}
3D models play an essential role in all areas of computer graphics, such as games, animated movies or virtual reality. 
To effectively showcase these models or to assess their quality, not only model parameters are important, such as geometry and material, but also the selection of optimal views is crucial. 
Optimal views should ensure that the model complexity is appropriately communicated, and relevant structures are visible. 
Many quality measures have been developed to aid in the automatic selection of optimal viewpoints on 3D models. 
The applications range from obtaining vantage points for capturing stills in architecture~\cite{He2017architecture}, to initial camera positioning for complex scene inspection~\cite{meuschke2017automatic,heinrich2016evaluating,song2014mesh,secord2011perceptual}, camera control~\cite{lino2015intuitive} and recommendations for scientific visualization~\cite{Yang2019volumevisualization}.
Most common viewpoint quality measures aim to measure the information content of rendered 2D images of 3D models.
The information content is usually derived from the visibility of the model geometry, making it sensitive to mesh quality and, in some cases, discretization.
For this reason view quality measures generally assume the geometry to be a clean watertight surface mesh, which is not always available in real world applications. Faulty meshing on the other hand, such as holes in the geometry or self-intersecting triangles, distort the resulting viewpoint quality~\cite{Bonaventura18}.
Finally, to find the best viewpoint existing work often renders the 3D model for a large set of candidate viewpoints, which makes finding optimal views a costly brute-force search~\cite{vazquez2002viewpoint,freitag2015comparison, kim2017category}. 

In this paper, we present the first optimal viewpoint learning approach, and demonstrate its applicability by learning different existing viewpoint qualities. 
To overcome the above mentioned limitations of finding optimal views, we train a deep neural network end-to-end to predict high quality viewpoints directly from the 3D model, dropping the rendered image from the optimization.
This makes our approach independent from rendering during evaluation, which means that the time-consuming rendering is separated from the optimal viewpoint prediction.
Hence, in contrast to previous work, our learned approach allows instantaneous predictions, making an expensive brute-force search over many rendered view alternatives unnecessary, and reducing prediction times from several minutes to a fraction of a second. 
In order to reduce the influence of mesh quality and discretization, we only use unstructured surface points as input, i.e., no information about the polygonization is provided to the network.
This forces the network to predict optimal views from an implicitly learned latent geometry representation, which, by design, is independent of the actual mesh polygonization. 
By training the network on clean meshes, we bias the latent representations towards clean surfaces, and as a result during evaluation the network will estimate a latent representation of a clean surface from the given points.
These considerations make our approach robust to the model’s discretization and mesh quality during evaluation, which enables the prediction of optimal viewpoints on a wide range of 3D models from different sources with varying mesh quality. 

However the end-to-end mapping from 3D model to optimal viewpoint is not well defined, as viewpoint quality measures do not necessarily have a unique maximum, but may have several, for instance, but not exclusively, due to model symmetries.
This ambiguity leads to conflicting ground truth information, resulting in opposing gradients which prevent meaningful learning. 
Existing techniques to resolve such label ambiguities only work for specific settings, e.g., local ambiguity~\cite{gao2017deep} or symmetric ambiguity~\cite{liao2019spherical}.
Thus we propose a more general approach, the dynamic label generation, which integrates the label decision into the training process.
This allows the network to dynamically adjust the labels during training which results in a harmonized label decision over the dataset, effectively reducing the influence of contradicting label decisions, and thus gradients, and enabling learning for this more general type of ambiguity.

\noindent Thus, within this paper we make the following contributions:

\begin{itemize}
    \item We present the first learning-based approach that directly predicts optimal viewpoints directly on 3D models, while being robust to the input mesh quality.
    \item We introduce a novel dynamic label generation method, incorporating the label decision into the training to resolve label ambiguity.
    \item We release viewpoint quality annotations for a subset of ModelNet40, which makes it the largest available viewpoint quality dataset -- by a large margin.
\end{itemize}

\section{Related Work}
\label{sec:Related Work}

The search for a good viewpoint of a 3D object is a problem that can be dated back to ancient societies such as the Greeks and Romans. Several rules such as the golden ratio, or the rule of thirds have been proposed to estimate beauty or proportion. More recently, the search for preferred views has also been addressed, especially in computer vision tasks (e.g., for object recognition), and researchers have wondered what parameters constitute a good view~\cite{Polonsky:2005:WIA}. Blanz et al. asked users about their preferences and dubbed preferred views of known objects \emph{canonical views}~\cite{blanz1999object}. They also found that in some cases, these correspond to three-quarter views (also with notable exceptions, such as in the case of vehicles). Secord et al. also analyzed viewpoint preferences in a large scale user study, and derived a combination of existing techniques~\cite{secord2011perceptual}. \\
Unfortunately, when developing an algorithm to find the best view, the orientation of the objects is commonly unknown, so for instance obtaining three-quarter views from loaded models cannot be done straightforwardly. Thus, algorithms tend to measure elements that are available through the geometry, such as triangles, silhouettes, depth maps, etc. 

\noindent\textbf{Viewpoint selection.}
The automatic selection of viewpoints for 3D scenes has many applications such as helping observers gain understanding on a certain scene~\cite{andujar2004way,liu2014flying,freitag2017assisted}, for object recognition~\cite{deinzer2006integrated,deinzer2009framework}, assisting in robotic tasks~\cite{saran2017view4robots}, inspection of volumetric models~\cite{viola2006importance,vazquez2008interactive,muhler2007viewpoint,yao2008intelligent,tao2009structure,meuschke2017automatic}, proteins~\cite{vazquez2002viewpoint,heinrich2016evaluating}, or scene reconstruction~\cite{marchand1999active,smith2018aerial}.
Depending on the task to be solved, the algorithms use the available data, sometimes only geometry (e.g.,~\cite{vazquez2002viewpoint,lee2005mesh}, and sometimes combined with user-defined importance (e.g.,~\cite{bordoloi2005view,muhler2007viewpoint}) to define viewpoint quality criteria. \\
Other researchers focus on combining multiple viewpoint qualities, e.g., with linear regression~\cite{kim2017category, secord2011perceptual}, to reflect the result of user studies.
Recently, also deep learning has been used in the creation of saliency maps~\cite{kim2017category}, to score candidate viewpoints~\cite{Yang2019volumevisualization} or to estimate viewpoint distributions of photographies~\cite{zhang20203d}.
Discussing the dozens of such techniques would be beyond the scope of this work, and thus we would like to guide the interested reader to some of the comparisons that have been published in literature~\cite{secord2011perceptual,freitag2015comparison,Bonaventura18}.

Despite the number of articles devoted to this issue, little has been done to generate fast algorithms for good viewpoint selection. In most cases, the measures require inspecting a very dense set of candidate views, which is time consuming. Accelerations presented in literature are typically greedy algorithms (e.g. for light source positioning~\cite{gumhold2002maximum}, or for volumetric models~\cite{monclus2012efficient}). Our learned viewpoint prediction outperforms all these methods by design, as one forward pass through the network enables viewpoint prediction in milliseconds, rather than minutes, which are required by the brute-force approaches.

\noindent\textbf{Label ambiguity.} 
Ambiguous labels are present in many tasks, such as image classification, image segmentation, pose-estimation or age estimation~\cite{gao2017deep} and can hurt the performance of a learner if not considered~\cite{Rupprecht2017MHP}.
There are different sources for these ambiguities, some tasks naturally allow multiple correct labels, e.g., in image classification an image can contain multiple objects, for other tasks it is difficult to provide a definitive label, e.g., it is hard to determine the exact pose of a partially occluded person.
While classification tasks can resolve label ambiguity to some degree by design, regression tasks often struggle with ambiguous label information. 
While restating a regression as a classification is possible~\cite{shi2019cnns}, it limits the possible performance by discretizing the output space. 
In cases where ambiguity exclusively stems from symmetry, partial restatement can be a trade-off, e.g., to resolve axial symmetry~\cite{liao2019spherical} or rotational symmetry~\cite{Corona2018rotationsymmetry}.
The problem of label ambiguity can also be viewed as a problem of contradicting gradients.
While the influence of such gradients can be reduced using mixtures of experts \cite{jacobs1991adaptive, jordan1994hierarchical}, where multiple experts are trained together with a gating network to divide the problem space into disjoint regions, each having its own expert, this method is not applicable to the problem of label ambiguity which is not separable in the input space, e.g., the same data point could be present twice in the dataset with different labels.

In contrast to these approaches, we present a novel dynamic label generation, which integrates the label generation into the training stage and harmonizes the label decision without further assumptions or restrictions.

\section{Viewpoint Quality Measures} \label{sec:vqs}
To demonstrate the proposed deep learning \revised{technique}, we have considered four different viewpoint quality measures\revised{, which we selected based on their effectiveness in previous studies and their popularity:}
Viewpoint Entropy (\VE)~\cite{vazquez2001viewpoint}, Visibility Ratio (\VR), also referred to as surface area~\cite{plemenos1996intelligent}, Viewpoint Kullback-Leibler divergence (\VKL)~\cite{sbert2005}, and Viewpoint Mutual Information (\VMI)~\cite{feixas2009unified}, which are defined as:

    \begin{alignat}{2}
        \mathrm{VE} &= -&&\sum_{z\in\mathcal{Z}} \frac{a_z(v)}{a_t(v)}\log\frac{a_z(v)}{a_t(v)},\\
        \mathrm{VR} &= &&\sum_{z\in\mathcal{Z}}vis_z(v)\frac{A_z}{A_t},\\
        \mathrm{VKL} &= &&\sum_{z\in\mathcal{Z}}\frac{a_z(v)}{a_t(v)}\log\frac{a_z(v)A_t}{a_t(v)A_z},\\
        \mathrm{VMI} &= &&\sum_{z\in\mathcal{Z}}p(z|v)\log \frac{p(z|v)}{p(z)},
    \end{alignat}
    where we follow the notation of Bonaventura et al.~\cite{Bonaventura18}:\
    \begin{align*}
        z           &\qquad     \text{polygon}\\
        \mathcal{Z} &\qquad     \text{set of polygons}\\
        vis_z(v)    &\qquad     \text{visibility of polygon } z \text{ from viewpoint } v \text{ (0 or 1)}\\
        a_z(v)      &\qquad     \text{projected area of polygon } z \text{ from viewpoint } v\\
        a_t(v)      &\qquad     \text{projected area of the model from viewpoint } v\\
        A_z         &\qquad     \text{area of polygon } z \\
        A_t         &\qquad     \text{total area of the model}\\
        p(z|v)      &\qquad      = a_z(v) / a_t(v),
                                \text{ conditional probability of } z \text{ given } v\\
        p(z)        &\qquad     \text{probability of } z
                                \text{ \revised{(average projected area of }} z)
    \end{align*}
    
    \revised{According to Bonaventura et al.~\cite{Bonaventura18}}, \VE\ and \VMI\ are the most popular viewpoint quality metrics used in most papers, 
    When evaluating them regarding user preference, Bonaventura et al. also found that these were in both extremes of the user preference spectrum. 
    While the views selected by \VE\ are highly preferred by users, the ones selected by \VMI\ were not always deemed as informative.
    \revised{Further, Secord et al.~\cite{secord2011perceptual} ranked \VR\ and \VE\ as the two most preferred. 
    Finally we added \VKL\ as it is partially sensitive to the models discretization, and the other ones were in the extremes (non-sensitive/highly sensitive).}
    Other measures could also be considered, however we restricted ourselves to these four measures as they represent the range of two main properties, 
    see Table~\ref{table_vq_properties}.
    
    
    \begin{table}[t]
        \centering
        \caption{\textbf{Viewpoint measure properties.}
            Two main properties, the correlation to user preference and the sensivity to mesh discretization, for the considered viewpoint measures~\cite{Bonaventura18}.}
        \begin{tabular}{lll}
            \toprule
             Measure & User Preference & Mesh Discretization \\
            \midrule
            \VE  & high & sensitive \\
            \VR & medium high & insensitive \\
            \VKL & medium low & near insensitive \\
            \VMI & low & insensitive \\
            \bottomrule
        \end{tabular}
        \label{table_vq_properties}
    \end{table}
    
The best viewpoints for \VR\ and \VE\ correspond to the highest viewpoint quality values, and for \VKL\ and \VMI\ to the lowest viewpoint quality values.
These viewpoint quality measures are defined for polygonal models and thus are, in contrast to our approach, dependent on the actual meshing with various degrees. 
While \VR\ and \VMI\ are insensitive to the discretization of the model, and \VKL\ is near insensitive, they all still assume clean surface meshes, as for example self-intersections of polygons change $A_t$ and $A_z$ and thus also \VR\ and \VKL\ , without necessarily altering the visible surface. 
This underlying assumption makes it harder to compare good viewpoints for models under different meshing qualities or resolutions, which is a problem if we want to extract model-spanning features of good viewpoints and bias the network towards good viewpoints of clean surface meshes.
We reduce these influences with a mesh cleaning pipeline (see Section \ref{sec:data_preproc}), in order to ensure that the viewpoint quality measures work as expected for different meshes.

To compute the best viewpoints for a given model we sample the unit sphere $\mathcal{S}^2$ with $1k$ viewpoints $\mathcal{V}\subset\mathcal{S}^2\subset\mathbb{R}^3$ on a Fibonacci sphere~\cite{gonzalez2010measurement}, generating almost equidistantly distributed viewpoints, on which we evaluate the four viewpoint quality measures. 
\revised{Compared to other work with 240~\cite{secord2011perceptual},  258~\cite{dutagaci2010benchmark} or 642~\cite{Bonaventura18} viewpoints, we achieve a denser sampling of the view sphere.}

\begin{figure}[t]
        \centering
        \input{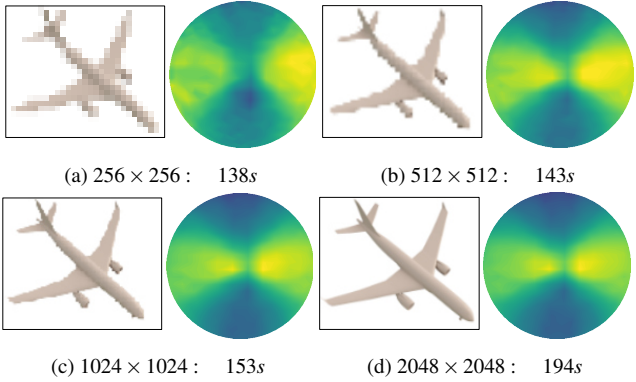}
            \caption{\textbf{Influence of image resolution.}
                Projection of \VE\ on the viewpoint sphere from $+ y$-axis for model \emph{airplane\_0275} from ModelNet40
                computed at different resolutions $256^2, 512^2, 1024^2, 2048^2$ 
                and the time needed to sample the \revised{$1k$ viewpoints} $\mathcal{V}$, averaged over 10 runs. 
                We choose a resolution of $1024^2$ as a trade-off between accuracy and speed. Note that the locations of the maxima (yellow) are stable at higher resolutions.}
            \label{fig:res_test}
    \end{figure}

\revised{The viewpoint quality measures are computed on rendered 2D images, and are thus influenced by the image resolution.} We compared different image resolutions, see Fig.~\ref{fig:res_test}, and chose to render the 3D models with $1024\times1024$ pixels, where the camera is placed at a distance of half the diagonal of the bounding box, centered on the mean of the bounding box, using perspective projection. 
We found this a good trade-off in accuracy and compute time. 

We further normalized each measure to the range $[0,1]$, where 0 and 1 refer to the viewpoint quality of the worst and best viewpoint, respectively:
\begin{align}
    VQ^*(v) = \frac{VQ(v)-VQ(v^-)}{VQ(v^+) - VQ(v^-)} ,
\end{align}
where $v^+\in\mathcal{V}$ is a viewpoint with the best and $v^-\in\mathcal{V}$ is one with the worst viewpoint quality of the sampled views $\mathcal{V}$. 
In the following we will always refer to these normalized versions of the viewpoint quality measures.
    
    \begin{figure*}[ht]
        \centering
        \includegraphics[width=\linewidth]{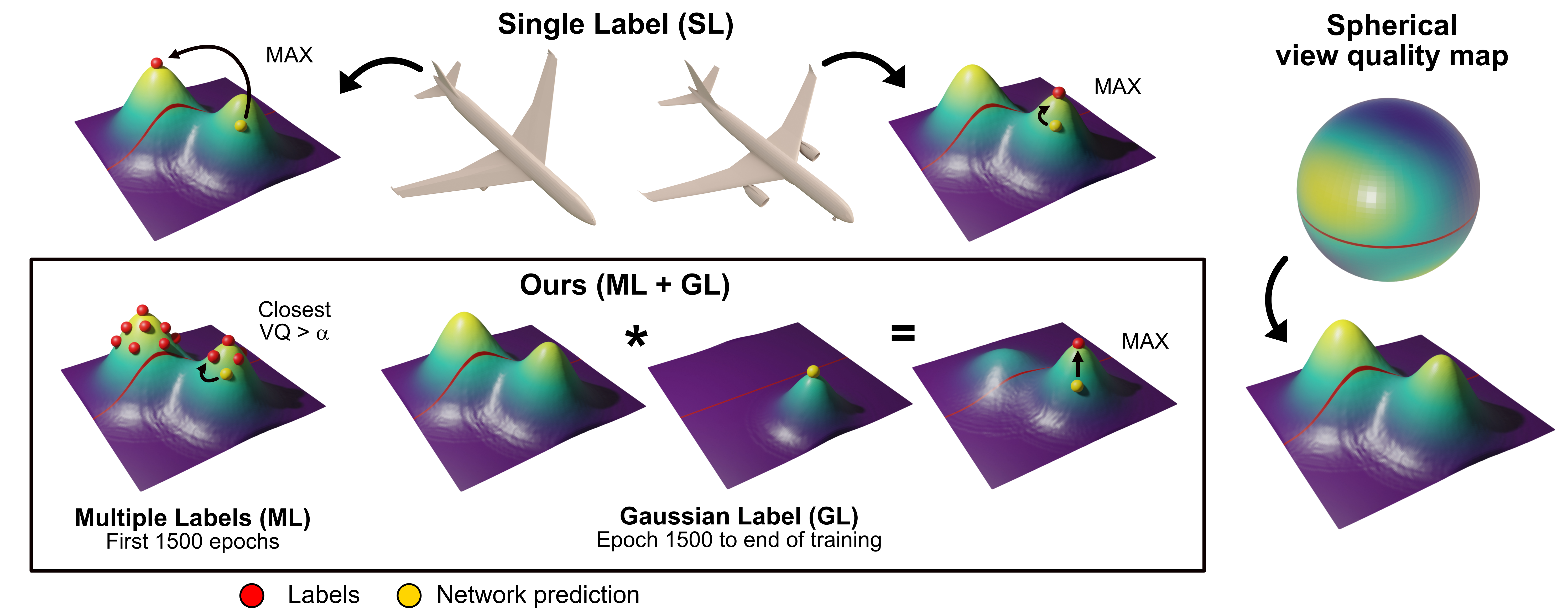}
        \caption{\textbf{Dynamic label generation.}
            Illustration of the proposed dynamic label generation technique for best viewpoint prediction, we use Mercator projections of the viewpoint sphere, as indicated on the right.
            \emph{Top: }
                Best viewpoints are not necessarily unique, thus randomly choosing a maximum as label can create different labels for similar input models, which the network is unable to resolve. 
            \emph{Bottom: }
                To harmonize the label decision, we propose our two stage dynamic label generation. 
                We first provide the network with multiple labels (ML) of high viewpoint quality and optimize towards the closest one. The labels typically form clusters in high quality areas, in which case the optimization tends to converge towards the boundary.
                To refine the predictions, we generate the label dynamically in a second stage (GL). The viewpoint quality distribution is weighted with a Gaussian centered at the current prediction and the maximum of the result is used as a label, which is typically a close local maximum, i.e., the maximum of the closest cluster.
                Both stages, ML and GL, provide more similar labels for similar input.
            }
        \label{fig:label_methods}
    \end{figure*}

\section{High Quality Viewpoint Prediction}\label{sec:best_view_pred} 
Predicting good viewpoints with neural networks confronts us with two major challenges, the non-uniqueness of the best viewpoint and the mesh dependency of the viewpoint qualities. 
In the following sections we will describe how we address these challenges.

\subsection{Dynamic Label Generation} \label{sec:label_generation}
Optimal viewpoints are not necessarily unique, e.g., due to model symmetry, which means that instead of a definite optimal viewpoint $v^+\in\mathcal{V}$ we typically find a set of viewpoints $\mathcal{V}^+\subset\mathcal{V}$ which maximize the viewpoint quality measure.
This phenomenon is referred to as label ambiguity.

In the general setting of label ambiguity a set of labels $\mathcal{Y}$ is given together with a quality measure $p:\mathcal{Y} \rightarrow [0, 1]$, which in our case is given by the normalized viewpoint qualities $p(v)= VQ^*(v): v\in \mathcal{V} \rightarrow [0,1]$.




The na\"ive label decision would be to ignore label ambiguity and choose one label $y^+\in\mathcal{Y}$ as the Single Label (SL) for each model prior to training and train to minimize the loss $\ell(\hat{y}, y^+)$,
between the prediction $\hat y$ and the chosen label. For viewpoints the natural choice is the cosine distance
\begin{align} 
    \ell(\hat{v, v^+}) &=1-\frac{\hat{v}\cdot v^+}{||\hat{v}||_2 ||v^+||_2}\\
    &= 1 - \hat{v}\cdot v^+,
\end{align}
between the prediction $\hat{v}$ and one viewpoint $v^+$ with a viewpoint quality of 1. (Note: $\ell_2$ norms are $1$ as we evaluate on the unit sphere.) 
However, if this decision is not consistent over the entire dataset, the network is unable to resolve the label ambiguity during training, e.g., if two similar models with similar viewpoint quality distributions are labeled differently, the networks receives contradicting gradients impacting the learning capability, as illustrated in Fig.~\ref{fig:label_methods} (\emph{top}).

We aim to resolve this problem by moving the label decision from a preprocessing step into the training process, by making it dependent on the current network prediction. 
This way the label decision is implicitly learned by the network, and can change dynamically during training to harmonize the label decisions over the dataset. 
In the following we propose two techniques for dynamic label generation.

\noindent\textbf{Multiple Labels (ML).}
We choose a subset of high quality labels
\begin{align}
    \mathcal{Y}^+ :=\{y\in \mathcal{Y}\ |\ p(y)\geq \alpha\},
\end{align}
with a quality threshold $0\leq \alpha\leq1$.
During training the loss between the current prediction $\hat{y}$ and the closest label in $\mathcal{Y}^+$ is minimized,
\begin{align}
    \ell_{ML}(\hat{y}) = \ell\big(\hat{y}, \underset{y\in\mathcal{Y}^+}{\mathrm{argmin\ }}(||y-\hat{y}||)\big).
\end{align}
In our setting of viewpoint prediction this simplifies to

\begin{align}
    \ell_{ML}(\hat{v}) = \min_{v\in\mathcal{V}^+} \left(1-\hat{v}\cdot v\right).
\end{align}
where we select labels $\mathcal{V}^+$ with a quality threshold $\alpha=0.99$.

In practice $\mathcal{V}^+$ often consists of clusters covering areas of good viewpoint quality values, which are similar for similar input models, causing the gradients to reinforce each other.
However, as the network only optimizes to the closest label, we observe it stopping at the boundary of one of these clusters, rather than moving towards its center (see Fig. \ref{fig:label_methods}), which results in non optimal values. 
To further improve the performance we propose a second approach which considers the quality measure and not just a quality threshold.
    
\noindent\textbf{Gaussian Labels (GL).} 
We propose to select labels with a high quality value $p$ in the proximity of the current network prediction.
We incorporate this through a locality constraint by weighting the label distribution with a shifted Gaussian function 
\begin{align}
    p_g(y,\hat{y}) &= p(y) \cdot \left(\exp{\frac{-\|y-\hat{y}\|_2}{2\sigma^2}} + s\right),
\end{align}
and then optimize towards a label which maximizes this measure
\begin{align}
    y^+_g(\hat{y}) &= \underset{y\in\mathcal{Y}}{\mathrm{argmax}\ }p_g(y,\hat{y}).
\end{align}
The additive term $s$ ensures that distant high quality labels are not dismissed, which keeps the network from getting stuck in larger regions with low $p_\mathcal{Y}$ values.
For our experiments we set $\sigma=2, s=1$, which leads to
\begin{align}
    VQ_g(v,\hat{v}) &= VQ^*(v) \cdot \left(\exp{\frac{-\|v-\hat{v}\|_2}{8}} + 1\right),\\
    v^+_g(\hat{v}) &= \underset{v\in\mathcal{V}}{\mathrm{argmax}\ }VQ_g(v,\hat{v}), \\
    \ell_{GL}(\hat{v}) &=   1 - \hat{v}\cdot v^+_g(\hat{v}).
\end{align}

We observe this approach to keep optimizing towards a local maximum of $VQ_g$, whereby the value of this local maximum can be in some cases sub-optimal, e.g., if the initial guess of the network is in a bad region. 

For best results we use ML for initialization to first optimize towards the closest high quality viewpoint, followed by GL to refine the predictions inside a promising region, see Fig.~\ref{fig:label_methods}.

    
    
\begin{figure*}[ht]
    \centering
        \includegraphics[width=\linewidth]{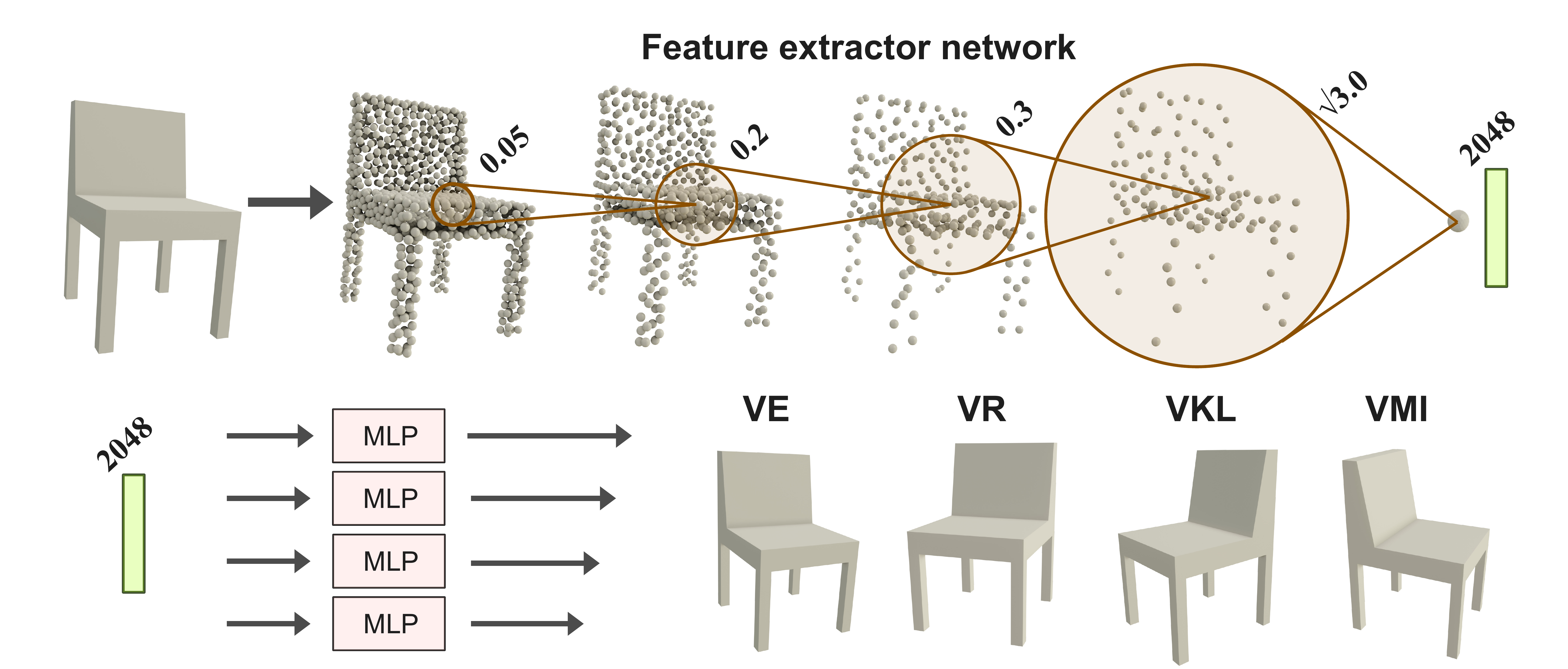}
    \caption{\textbf{Network architecture.}
        \emph{Top:}
        We use a MCCNN feature extractor, each layer performs a spatial convolution with increasing radius which increases the feature dimension 
        and reduces the spatial resolution, resulting in a $2048$ dimensional latent geometry representation.
        \emph{Bottom:}
        The latent representation is processed by parallel MLPs, each predicting the viewpoint for a different viewpoint quality measure.
        }
    \label{fig:net_architecture}
\end{figure*}

\subsection{Mesh Cleaning} \label{sec:data_preproc}
As mentioned above, some viewpoint quality measures are sensitive to the meshing of the models, and bad mesh quality can lead to distortions in the viewpoint quality computation. 
These inaccuracies reduce the comparability between different models, which makes it hard for a network to determine the important features. 
Providing clean and comparably meshed models on the other hand biases the network to implicitly extract features of a clean surface solely from point information.
To minimize these influences, we pass all meshes through a mesh cleaning pipeline, which resolves mesh intersections and regularizes the meshing. 
For details on the mesh cleaning and its influence on the viewpoint quality measures we refer the user to the supplementary material.
We note that our pipeline does not remove all artifacts but the achieved mesh quality proved sufficient for our experiments.

\subsection{Network Architecture} \label{sec:net_arch}
\revised{We deliberately chose point clouds as input to achieve robustness / independence to mesh polygonization / discretization, in contrast to neural networks which operate directly on meshes.
Further, we chose Monte-Carlo Convolutional Neural Networks (MCCNN)~\cite{hermosilla2018monte} over other point cloud architectures because of its robustness to input sampling by considering the point cloud density.}

Our feature extraction network consists of four convolutional layers with convolutional radii $0.05,0.2,0.3,\sqrt{3}$, relative to the bounding box of the model. Each convolutional layer operates on different resolutions, which are computed using Poisson disk sampling with radii $0,0.025,0.1,0.4,\sqrt{3}$, again relative to the bounding box of the model. 
The respective feature dimensions are $3,128,256,1024,2048$.
The feature extractor architecture is shown in Fig.~\ref{fig:net_architecture}~(\emph{top}). 
The learned latent representation is processed by four parallel Multi Layer Perceptrons (MLPs) with three layers of sizes $1024,256,3$, each outputting a viewpoint $\hat{v}\in\mathbb{R}^3$ for one of the four viewpoint quality measures, see Fig.~\ref{fig:net_architecture}~(\emph{bottom}).
We found that training one feature extractor for all four viewpoint qualities improves the performance as compared to training four separate networks.
An effect we account to the different losses improving the feature extractor, similar to auxiliary losses~\cite{Szegedy2015googlenet}.
We further use batch normalization and ReLU activation between all layers.

For all conducted experiments, we used the same hyperparameters, stressing that our network is applicable to different categories and viewpoint quality measures without further tuning. 
Namely we use dropout~\cite{JMLR:v15:srivastava14a} in the MLP layers with a dropout rate of $0.5$, Adam optimization~\cite{kingma2014ADAM} with batch size of $8$ and a learning rate decay with an initial learning rate of $0.001$ which is multiplied by $0.75$ every $200$ epochs.
We train for a total of $3000$ epochs and switch from ML to GL after $1500$. 
\section{Experiments}\label{sec:experiments}


To validate our viewpoint learning approach, which is enabled by dynamic label generation, we conducted three experiments. 
First, we trained a neural network to predict good viewpoints on point clouds of arbitrarily oriented 3D models, while we compare our label generation method to existing techniques.
Next we inspected the robustness of our method towards different meshing and samplings of the input models, and lastly provide timings for both the sampling algorithm and our network.

\subsection{Data}\label{sec:data}
All experiments were conducted on a subset of ModelNet40~\cite{wu20153dModelNet}, composed of the categories \emph{airplane, bench, bottle, car, chair, sofa, table} and \emph{toilet}, which we split into 80\% training, 10\% validation and 10\% test data. 
All models were preprocessed as described in Section~\ref{sec:data_preproc}. 
In order to sample the viewpoint quality measures in reasonable time we only use models with at most 10k faces. 
All meshes are converted into point clouds by sampling 20k random uniform points on the faces.
\revised{We use a rather dense input of 4096 points per model to capture fine geometric detail, for comparison common object classification networks typically use only 1024 points~\cite{hermosilla2018monte, qi2017pointnetplusplus}.}

\noindent\textbf{Data Augmentation}
Neural networks working with three dimensional input data usually require a large database to achieve noteworthy performance. 
This is due to the high dimensionality of the input space, as well as to the complexity of the tasks.
As the available sources for 3D data are rather limited, as compared for example to image data, the use of data augmentation, which increases the dataset virtually, are crucial for our experiments.

Therefore, we use the following two data augmentation strategies, random sampling and rotations:\\
1. The input point clouds are generated by selecting $1024$ points using farthest point sampling~\cite{eldar1997farthest}, and selecting additional $3072$ random points.\\  
2. The input point clouds are augmented with rotations from $\mathcal{SO}(3)$, whereby the three angles are chosen from a random uniform distribution on $[0,2\pi]$.
    \begin{figure*}[t]
        \centering
        \includegraphics[width=\linewidth]{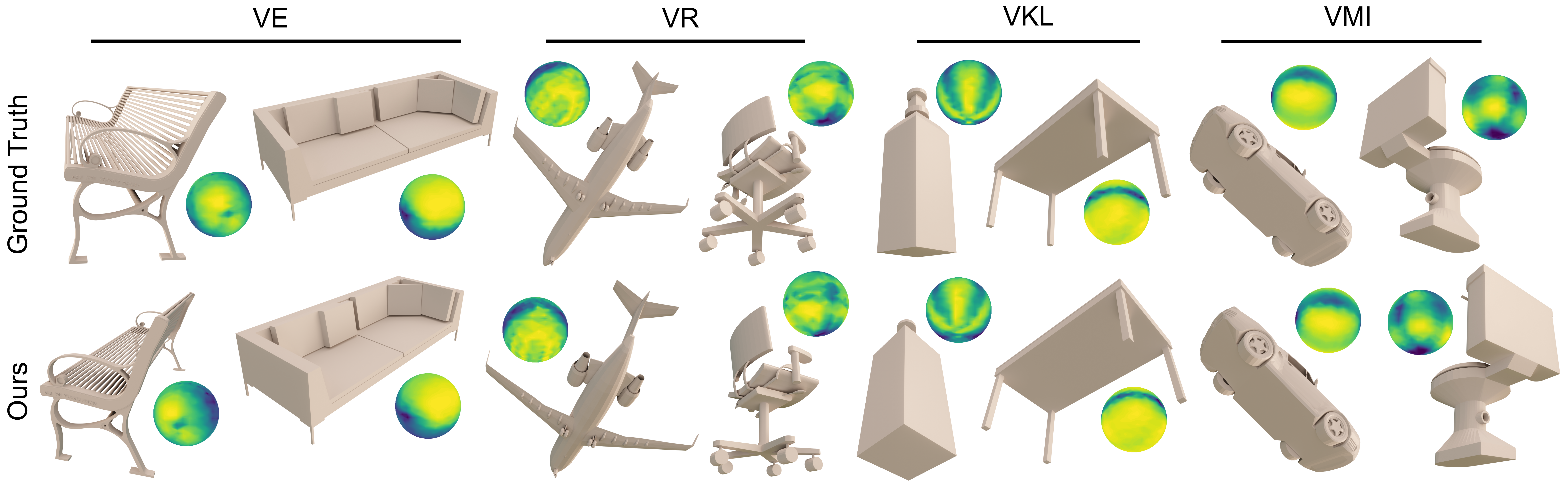}
        \caption{\textbf{Viewpoint prediction results for different viewpoint qualities.}
            Viewpoints predicted by our network for unseen models and ground truth achieved from sampling the viewpoint sphere.
            We also show the corresponding viewpoint quality spheres centered at the displayed viewpoint.
            The network successfully predicts high quality viewpoints, indicated by the yellow areas in the viewpoint spheres.
            }
        \label{fig:vqprediction}
    \end{figure*}

\subsection{Viewpoint Prediction} \label{sec:vq_experiment}

We demonstrate the effectiveness of our two stage dynamic label generation (ML+GL) by comparing it against single label cosine-distance (SL) and existing work on resolving label ambiguity, Deep Label Distribution Learning (DLDL)~\cite{gao2017deep}, to directly predict the viewpoint quality distribution, and Spherical Regression (SR)~\cite{liao2019spherical}, which splits the optimization into two parts, a regression for the absolute value $|v|$ and a classification task for the signs.

We train a network as described in Section~\ref{sec:net_arch} on each category to predict viewpoints for the four different viewpoint quality measures, the Viewpoint Entropy (\VE), the Visibility Ratio (\VR), the Viewpoint Kullback-Leibler divergence (\VKL) and the Viewpoint Mutual information (\VMI).
For SR and DLDL we have to adapt the architecture and loss as follows:

\noindent\textbf{SR}:
The loss function $\ell$ for SR consists of the cosine distance to $|v^+|$ and the cross entropy-loss of the sign prediction. 
Furthermore we use two MLPs per output to predict absolute values and the sign categories.

\noindent\textbf{DLDL}:
The loss function for DLDL is the per-pixel $\ell_2$ distance.
The MLPs are replaced by 2D decoder networks consisting of 2D deconvolutions and residual blocks~\cite{he2016deep} predicting the viewpoint quality distributions, for more details on the architecture we refer to the supplementary material. 
\revised{For a fair comparison we predict at a resolution of $32\times32=1024$, which is close to the $1000$ sampled viewpoints used as labels.}
We choose the viewpoint with the highest predicted viewpoint quality as the predicted viewpoint.

    \begin{table}[b]
        \centering
        \caption{\textbf{Viewpoint prediction results.}
            \emph{Top: }
            Mean viewpoint quality in \% of the predicted viewpoints using the different labeling techniques on the test set. 
            Using dynamically generated labels (ML, GL, ML+GL) improves over one stage methods (ML, GL) and existing methods (SL, DLDL, SR), where 
            our proposed two stage dynamic label generation method ML+GL yields best results for all four viewpoint quality measures.
            \emph{Bottom: } Mean viewpoint quality in \% of the ML+GL approach for the different categories.
                The performance is consistent over all categories.}
\begin{tabular}{llrrrr}
    \toprule
    labels & categories & \VE & \VR & \VKL & \VMI
    \\
    \midrule
        SL & mean & 62.4 & 71.0& 80.7  & 83.0 
        \\
        SR & mean & 63.1 & 69.8 & 80.6 & 80.1 
        \\
        DLDL & mean & 58.7 & 66.6 & 77.9 & 77.9 
        \\
        \textbf{Ours (ML+GL)} & mean & \textbf{79.3} & \textbf{78.2} & \textbf{91.2} & \textbf{92.5} 
        \\        
        Abl. 1 (ML only) & mean & 70.1 & 72.1 & 82.6 & 82.1 
        \\
        Abl. 2 (GL only) & mean & 74.2  & 75.1 & 89.3& 87.7 
        \\
    \midrule
        Ours (ML+GL) & airplane     & 79.1 & 74.8 & 95.2 & 96.6 
        \\
        Ours (ML+GL) & bench        & 67.7 & 72.8 & 85.5 & 87.3 
        \\
        Ours (ML+GL) & bottle       & 75.3& 78.0  & 94.9 & 94.1 
        \\
        Ours (ML+GL) & car         & 84.0 & 80.3 & 89.7 & 92.2 
        \\
        Ours (ML+GL) & chair       & 73.0 & 77.9 & 90.8 & 93.0 
        \\
        Ours (ML+GL) & sofa        & 88.8 & 75.7 & 92.2 & 93.5 
        \\
        Ours (ML+GL) & table       & 83.0& 82.0  & 91.6 & 90.1 
        \\
        Ours (ML+GL) & toilet      & 83.8 & 84.3 & 89.8 & 93.4 
        \\
    \bottomrule
\end{tabular}
        \label{table_vq_label_comparison}
    \end{table}

We measured the mean viewpoint qualities of the predicted viewpoints on the test set, averaged over all categories, and compared the different methods in Table~\ref{table_vq_label_comparison}. 
Our proposed two stage combination of ML and GL (ML+GL) clearly outperforms the na\"ive approach SL and both DLDL and SR.
Further, we performed an ablation study where we compare our combined ML+GL approach to only using multiple labels (ML) and Gaussian labels (GL).
While all three dynamic label generation methods outperform the existing methods with precomputed labels, confirming that our proposed method provides a better way to resolve label ambiguity for this task, the two stage ML+GL method improves over both single stage methods.
We conclude that initializing the predictions with ML substantially improves the results over training solely with GL, as GL has a stronger locality restriction, making it sensitive to initialization.

\begin{figure}[t]
        \centering
        \includegraphics[width=.8\linewidth]{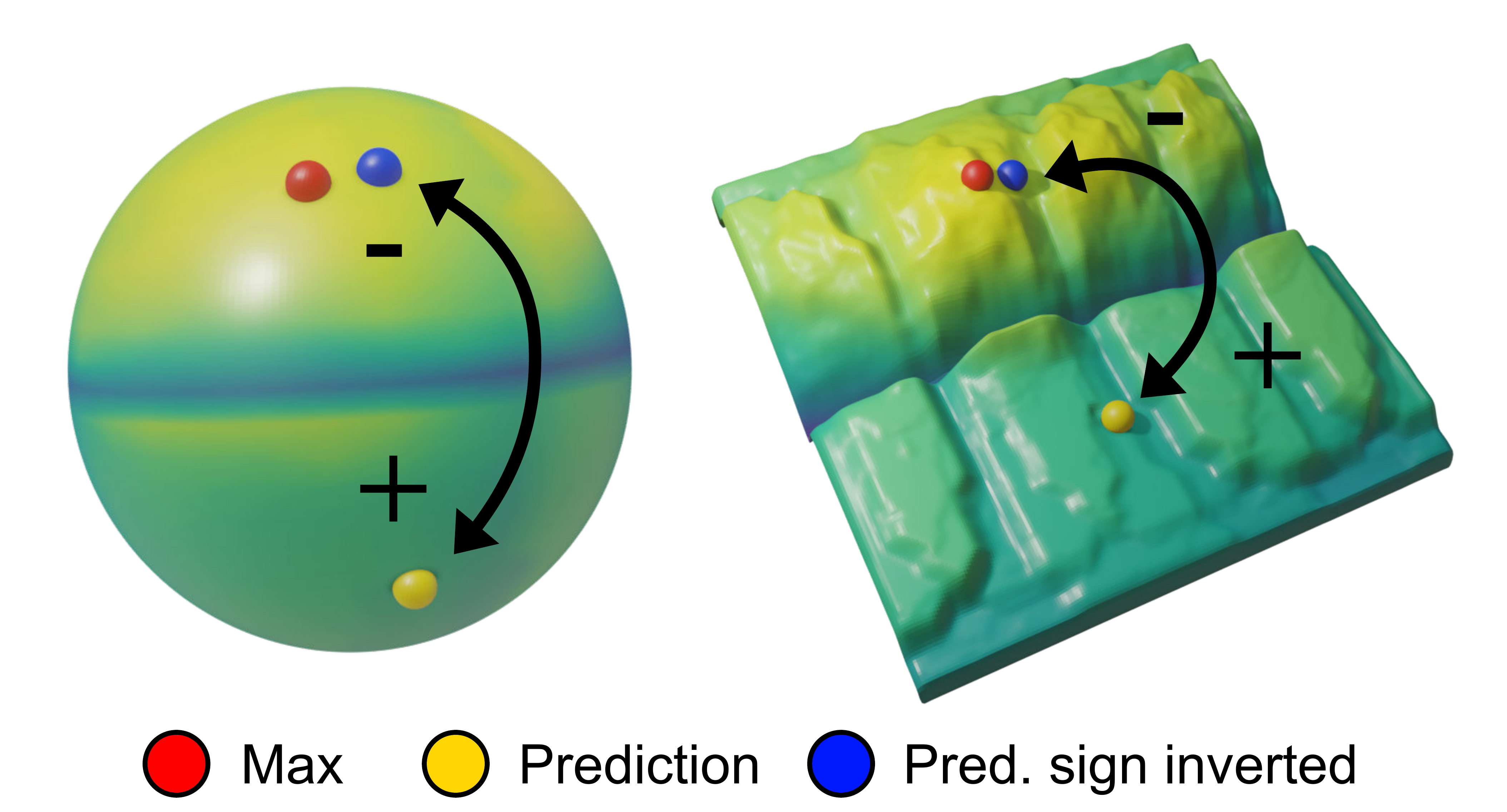}
        \caption{\textbf{Spherical regression.} SR struggles with resolving label ambiguity as the ambiguity is not axis-symmetric leading to predictions that have flipped sign decision (yellow) although the absolute value might be correct (blue).}
        \label{fig:sr_figure}
    \end{figure}

We found that SR is not always able to resolve the ambiguity leading to predictions with wrong sign decisions or false regression results for $|v|$, interpolating good viewpoints, see Fig.~\ref{fig:sr_figure}.
    \begin{figure}[b]
        \centering
        \includegraphics[width=\linewidth]{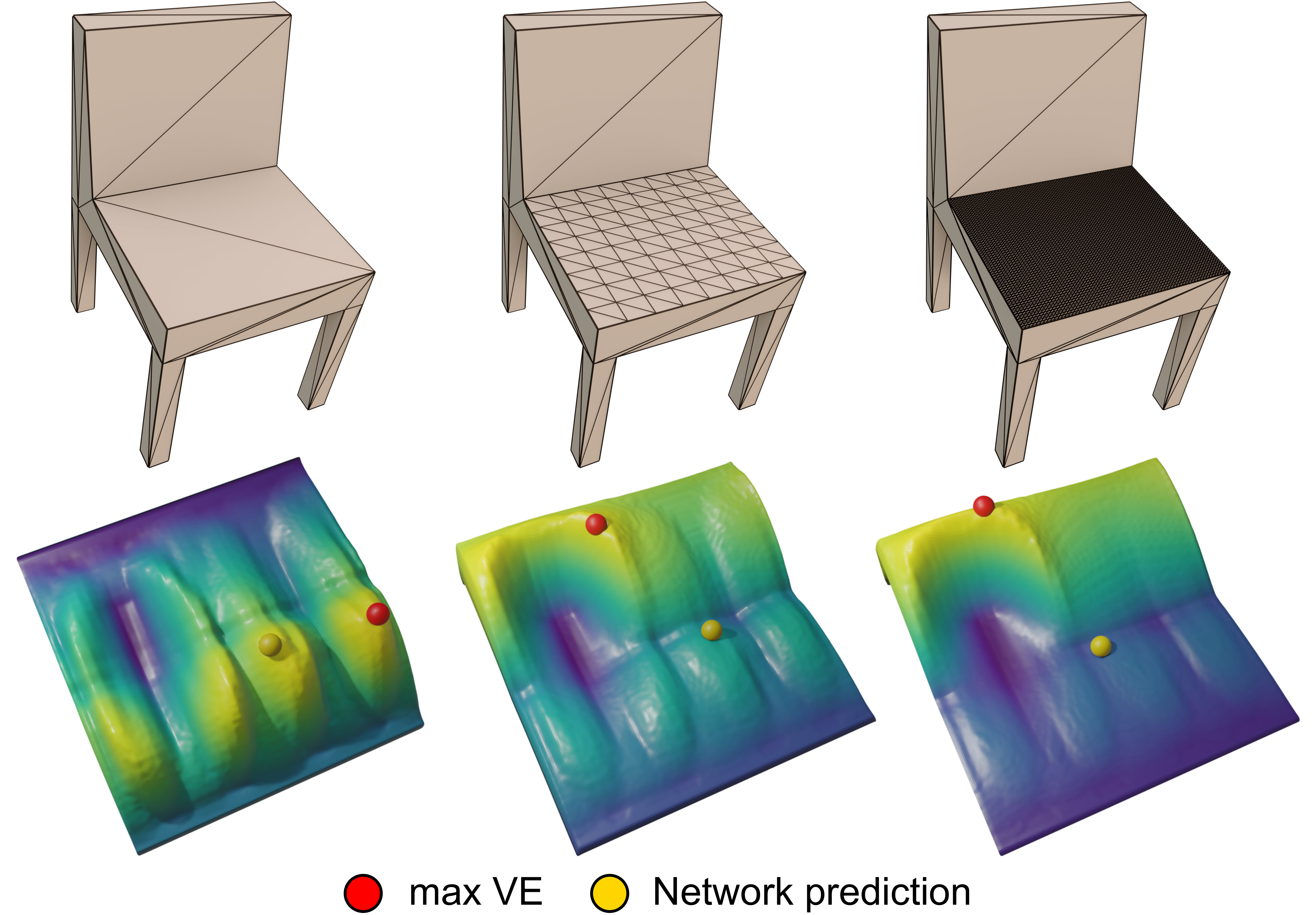}
        \caption{\textbf{Robustness to mesh polygonization.}
            We show predictions using \VE\ for different subdivisions of the seating surface.
            As \VE\ favors small triangles the bias towards views from the top increases with higher mesh density (red).
            Our network based approach remains stable independent of the meshing (yellow).}
        \label{fig:mesh_independence}
    \end{figure}
We theorize that this is because an underlying assumption for SR is that $|v|$ is the same for all labels, but as in our case the label ambiguity does not solely stem from model symmetry and the input is not necessarily aligned with the 3D axes, the assumption does not hold.

Predicting the viewpoint quality distribution (DLDL) resulted in the worst results.
By analyzing the network prediction we found that the predicted distributions are much smoother than the ground truth distributions, for details we refer the reader to the supplementary material. 
We  hypothesize that the network is unable to capture the geometric details which create the high frequency properties of the viewpoint quality distribution and as a result predicts an averaged distribution for similar models. 
We account this to two main factors.
First the tasks itself is harder than only predicting the optimal viewpoint which demands the extraction of geometric features at a finer scale. 
The extraction of such details would however require a denser input sampling and a wider and deeper feature extractor, and in consequence also a larger data set for training. 
Second the influence of mesh quality on the viewpoint quality distribution is naturally higher than on the position of the optimal viewpoint. 
Thus our preprocessing pipeline might be insufficient and create distortions that the network is unable to resolve.

The results of our method are stable for all examined categories as can be seen in the bottom half of Table~\ref{table_vq_label_comparison}, showing that no additional tuning of the hyperparameters is necessary to learn various categories or viewpoint quality measures, detailed results can be found in the supplementary material. 

Viewpoints predicted on the test set, i.e. unseen models, by our network trained with ML+GL labels can be seen in Fig.~\ref{fig:vqprediction}. 
We stress that due to label ambiguity the network is not optimized towards reproducing the same viewpoint as the sampling method, but to predict a viewpoint with high viewpoint quality.  
This potentially leads to different views, e.g., the toilet in Fig.~\ref{fig:vqprediction}, for which both views have a high quality, as can be seen in the viewpoint quality spheres in the figure.
    \begin{table}[b]
        \centering
        \caption{\textbf{Robustness to input sampling.}
            Comparison of the network performance \revised{at test time} for different input data: \emph{preprocessed}: by our mesh cleaning pipeline, \emph{raw:} from unaltered ModelNet40, \emph{surface}: point sampling of Qi et al.~\cite{qi2017pointnetplusplus}.
            The network achieves comparable results under different input meshing and point sampling methods\revised{, without further training.}}
        \begin{tabular}{lrrrr}
    \toprule
    source   & \VE  & \VR & \VKL & \VMI 
    \\
    \midrule
    preprocessed meshes        & 79.3 & 78.2 & 91.2 &  92.5 
    \\
    raw meshes             &  78.9 & 76.2 &  88.6  &  90.4 
    \\
    surface sampling        &  79.1 & 74.6 & 88.9 & 88.4 
    \\
    \bottomrule
\end{tabular}
        \label{tab:mesh_comparison}
    \end{table}    
\subsection{Mesh and Sampling Independence}\label{sec:mesh_independence}
We use unstructured 3D convolutions and hence the input to the network are point clouds only consisting of coordinate information. 
As these points carry no additional information about the polygonization of the underlying mesh we expect our approach to be insensitive to the discretization of the mesh \revised{at test time}. 
\revised{Furthermore, the use of MCCNNs, which consider an estimate of the point density, should result in a robustness to point sampling strategies.}

To confirm this we perform two different experiments.
The first one is the application to a toy example, in which we subdivide a part of the \emph{chair\_0047} mesh into smaller polygons.
On the original model \VE~ prefers views from the bottom showing more geometric details in form of the legs, while after subdividing the seating surface \VE~ mistakes the small faces as surface details, emphasizing the visibility of this area, see Fig.~\ref{fig:mesh_independence}, \revised{which results in a viewpoint far from the optimal views of the original mesh.}
Our approach on the other hand, predicts viewpoints in an optimal area of the original mesh, independent of the meshing.
    
\revised{In a second experiment we show the robustness of our approach in practice to input that differs from the clean data provided during training.
First, to investigate the robustness to mesh quality, we tested our network on the raw ModelNet40 models, which contain self-intersections, non-surface faces and non-uniform discretization.
This is a more challenging task than the first toy example, as the model geometries are different and not only the mesh discretization, which confronts the network with out-of-domain input.
Second, to show that our approach is additionally robust to different point sampling strategies,} we also evaluate on the points provided by Qi et al.~\cite{qi2017pointnetplusplus}, who use a different pipeline to achieve clean surface point clouds.
The results reported in Table~\ref{tab:mesh_comparison} confirm that our approach is robust under sampling of the input data.
We infer that the network has learned an internal representation of the meshing used during training
.
    
\subsection{Timings}\label{sec:timings}
We compared the time needed to estimate high quality viewpoints using the sampling approach described in Section~\ref{sec:vqs}, and the time needed to predict high quality views using our neural network model, as described in Section \ref{sec:net_arch}. 
The timings were measured on a system with an Intel Core i7-8700K CPU @ 3.70GHz and a NVIDIA GeForce GTX 1080 GPU. While the sampling approach was implemented using Python and OpenGL, our network approach was realized through Python and TensorFlow. 
To make the measurements comparable, we employed the following two conditions. First, we neglected initialization times, which include loading the meshes, preprocessing the meshes for the sampling method and sampling points and loading the weights for the network. 
Second, we sampled the viewpoint quality measures in one procedure, computing shared values only once.
For the evaluation we chose models of different sizes, ranging from 10k faces to 1M faces, whereby we processed all these models 10 times with both methods and reported the averaged times in Table~\ref{tab:timings}.

While the elapsed time of the sampling approach is approximately linear in the number of candidate views and the number of faces the network only requires one execution. 
This execution's time is independent of the model size, outperforming the other method in orders of magnitude.
While we see some variation in the execution time of the network, which we account to varying numbers of points in the 3D convolutions and point hierarchy levels, the timings are comparable for all inspected models.
        \begin{table}
            \centering
            \caption{\textbf{Time comparison.}
                Elapsed time of sampling based methods and ours for different model sizes, all timings are averaged over 10 executions.
                We measure the brute force sampling method using $250$, $500$ and $1k$ candidate views, and measure our model when batch processing $1$, $64$ and $256$ models at the same time.
                Our network approach is faster in orders of magnitude and is independent of the model size as it uses a point cloud of fixed size. 
                We report N/A where the execution did not finish after $12h$.
                }
            



\begin{tabular}{r|rrr|rrr}
    \toprule
    & \multicolumn{3}{c}{sampling}  & \multicolumn{3}{c}{ours} 
    \\
    \midrule
    & \multicolumn{3}{c}{number of views}  & \multicolumn{3}{c}{batch size}
    \\
     \#faces & 250 & 500 & 1000 & 1 & 64 & 256 
    \\
    \midrule
    10k & 20s & 40s  & 79s  & 0.263s  & 0.015s  & 0.012s
    \\
    50k & 92s & 184s  &  373s & 0.253s  & 0.013s & 0.010s
    \\
    100k & 178s  & 356s  & 722s  & 0.260s  & 0.018s  & 0.015s
    \\
    400k & 737s  & 1479s  & 2929s  &  0.270s & 0.020s  & 0.017s
    \\
    1M &  2030s & N/A  & N/A  & 0.258s  & 0.010s  & 0.007s
    \\
    \bottomrule
\end{tabular}
            \label{tab:timings}
        \end{table}
\section{Limitations \revised{and Future Work}}\label{sec:limitations}
To achieve the reported results, we trained category specific instances of our network in a divide-and-conquer scheme, which is common for similar deep learning tasks such as viewpoint estimation~\cite{shi2019cnns} or upright prediction~\cite{liu2016upright}.
This prevents the proposed network from generalizing to unseen categories,
\revised{however, we see no theoretical limitation of our method and expect such generalization to be possible in the future by i) expanding the learning capabilities, e.g. using mixture of experts as was shown for viewpoint estimation~\cite{liao2019spherical}, and ii) increased amount of training data, a key ingredient in order to generalize to unseen categories.}

\revised{While our network can predict multiple viewpoints at once, the views are independent, as it predicts one viewpoint per measure.
We see potential for predicting multiple viewpoints that compliment one another.
However, this leads to the problem of defining a good second view. 
Is it one that best covers the unseen parts of the model or a second view with high quality value? 
Note that the latter can be a very similar view direction.
Moreover, the number of good views may vary per model, which could be addressed with network architectures that can output sequences, e.g. recurrent models.}

\revised{Our method learns good viewpoints based on existing viewpoint quality measures, however, no measure is able to fully model human preference.
While our method is general enough to learn on manually selected viewpoints,
currently no large scale data set is available, and existing data is too limited for deep learning (16 models~\cite{secord2011perceptual}, 68 models~\cite{dutagaci2010benchmark}).
}
\revised{A way to overcome the need for a large data set would be self- or weakly-supervised training, which could in future be investigated based on recent advances in differentiable rendering~\cite{nimier2019mitsuba, nguyen2018rendernet}.}

\revised{Furthermore, we see potential to induce non-geometrical biases to the network by considering semantics, e.g. up-right orientation.}
    
\section{Conclusion}\label{sec:conclusion}
The proposed learned viewpoint prediction provides a way to predict high quality viewpoints for different viewpoint quality measures and model categories.
By separating viewpoint selection and rendering our approach performs faster than existing techniques by several orders of magnitude. 
\revised{This makes our method suitable for applications which benefit from speed and parallelizability, such as automatic thumbnail generation of 3D data sets or initial camera placement for user interaction.}
The prediction of viewpoints directly from unstructured 3D point data proved to make the prediction robust to meshing properties, which makes us believe that the network has learned an internal representation of a clean mesh, as intended.
The proposed dynamic label generation method is essential to resolve label ambiguity during training, outperforming existing methods, and is designed to be transferable to other learning tasks that involve label ambiguity.

On top of the contributions made in this article, we provide a dataset, which will be, to our knowledge, the first large scale viewpoint quality dataset containing more than 16k models in total, more details can be found in the supplementary material.  
    


\bibliographystyle{eg-alpha-doi}

\bibliography{_bib}


\appendix 
    \begin{table*}
        \begin{center}
            \caption{\textbf{Detailed results.} Breakdown from  Table 2 for each category. 
}
\begin{tabular}{llrrrrrrrrr}
    \toprule
     & & airplane & bench & bottle & car & chair & sofa & table & toilet & \hspace{3mm}mean
     \\
     \midrule
    \multirow{5}{*}{\VE}
        & SL & 60.7 &  50.9 &  64.8 &  58.0 &  63.8 &  88.7 &  38.3 &  74.4 &  62.4 
        \\
        & SR & 49.4 &  65.1 &  53.3 &  64.0 &  66.3 &  63.4 &  73.5 &  70.0 &  63.1 
        \\
        & DLDL & 47.0 &  55.4 &  53.4 &  58.7 &  62.9 &  63.4 &  56.5 & 72.4 &  58.7
        \\
        & ML & 55.4 &  62.1 &  50.4 &  79.8 &  \textbf{73.9} &  83.3 &  76.4 &  79.2 &  70.1 
        \\
        & GL & 70.0 &  \textbf{69.5} &  52.9 &  82.0 &  71.7 &  87.9 &  76.6 &  83.4 &  74.2 
        \\
        & ML+GL & \textbf{79.1} &  67.7 &  \textbf{75.3} &  \textbf{84.0} &  73.0 &  \textbf{88.8} &  \textbf{83.0} &  \textbf{83.8} &  \textbf{79.3} 
        \\
    \cmidrule{2-11}
    \multirow{5}{*}{\VR}
        & SL & 71.4 &  71.7 &  69.9 &  65.1 &  72.4 &  \textbf{76.4} &  76.4 &  64.9 &  71.0 
        \\
        & SR & 71.0 &  73.4 &  72.2 &  69.3 &  65.3 &  59.2 &  80.3 &  67.4 &  69.8 
        \\
        & DLDL & 69.5 &  70.2 &  72.6 &  66.3 &  67.0 &  59.5 &  69.3 & 58.9 &  66.6
        \\
        & ML & 63.2 &  73.6 &  70.0 &  74.2 &  77.5 &  73.3 &  81.0 &  64.0 &  72.1 
        \\
        & GL & 66.2 &  \textbf{83.0} &  69.1 &  78.6 &  75.5 &  75.5 &  80.8 &  72.4 &  75.1 
        \\
        & ML+GL & \textbf{74.8} &  72.8 &  \textbf{78.0} &  \textbf{80.3} &  \textbf{77.9} &  75.7 &  \textbf{82.0} &  \textbf{84.3} &  \textbf{78.2} 
        \\
    \cmidrule{2-11}
    \multirow{5}{*}{\VKL}
        & SL & 89.2 &  76.2 &  74.9 &  83.7 &  83.5 &  86.3 &  86.0 &  65.6 &  80.7 
        \\
        & SR & 86.2 &  84.4 &  88.9 &  74.0 &  72.1 &  79.1 &  89.0 &  71.5 &  80.6
        \\
        & DLDL & 86.7 &  83.3 &  88.9 &  73.0 &  72.0 &  73.0 &  78.8 & 67.8 &  77.9
        \\
        & ML & 79.7 &  79.8 &  92.7 &  80.4 &  86.2 &  75.5 &  90.2 &  76.5 &  82.6 
        \\
        & GL & 91.8 &  \textbf{88.1} &  90.9 &  85.3 &  89.3 &  \textbf{94.0} &  90.9 &  84.4 &  89.3 
        \\
        & ML+GL & \textbf{95.2} &  85.5 &  \textbf{94.9} &  \textbf{89.7} &  \textbf{90.8} &  92.2 &  \textbf{91.6} &  \textbf{89.8} &  \textbf{91.2} 
        \\
    \cmidrule{2-11}
    \multirow{5}{*}{\VMI}
        & SL & 90.6 &  79.2 &  80.4 &  84.0 &  88.3 &  92.6 &  84.3 &  64.4 &  83.0 
        \\
        & SR & 85.0 &  86.2 &  86.5 &  81.6 &  70.6 &  75.7 &  \textbf{91.1} &  64.0 &  80.1 
        \\
        & DLDL & 87.8 &  79.8 &  86.9 &  79.6 &  72.1 &  69.9 &  80.0 & 67.0 &  77.9 
        \\
        & ML & 88.7 &  68.7 &  92.1 &  80.2 &  89.9 &  81.9 &  87.7 &  67.8 &  82.1 
        \\
        & GL & 94.0 &  85.1 &  91.5 &  78.7 &  91.0 &  91.2 &  88.8 &  81.7 &  87.7 
        \\
        & ML+GL & \textbf{96.6} &  \textbf{87.3} &  \textbf{94.1} &  \textbf{92.2} &  \textbf{93.0} &  \textbf{93.5} &  90.1 &  \textbf{93.4} &  \textbf{92.5} 
        \\
    \bottomrule
\end{tabular}

            \label{table_vq_results_detail}
        \end{center}
    \end{table*}
\section{Mesh Cleaning Pipeline}\label{app:mesh_cleaning}    
    \begin{figure}[b]
        \centering
        \includegraphics[width=\linewidth]{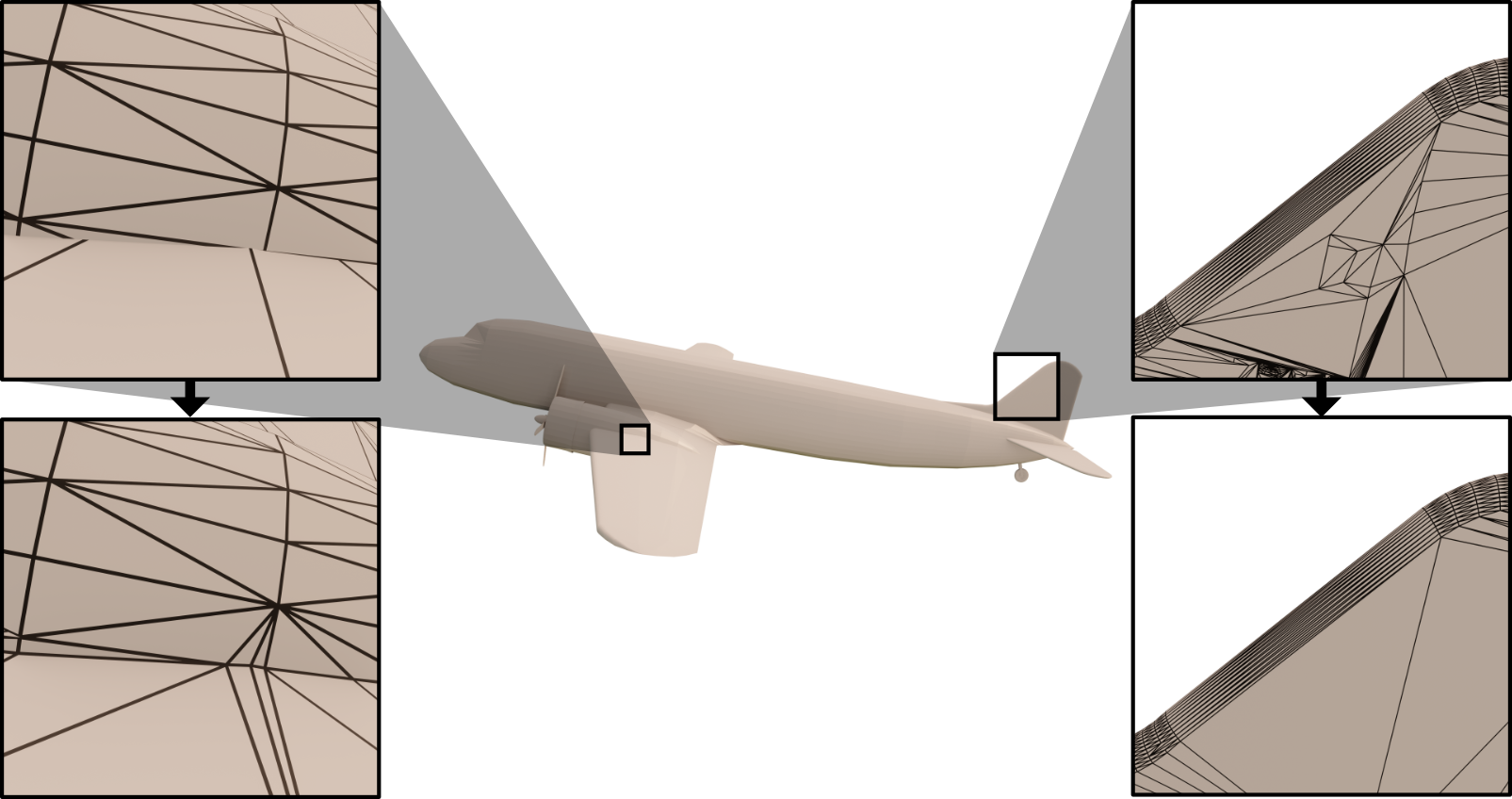}
        \caption{\textbf{Mesh cleaning.}
            Results of the different steps to clean the meshes on \emph{airplane\_0004}. 
            The original mesh contains self intersections (left) and non-uniform meshing artifacts (right), which are resolved in the first and third step our mesh cleaning pipeline, respectively.}
        \label{fig:mesh_cleaning}
    \end{figure}
    Our mesh cleaning pipeline consists of three steps.
    First we resolve self intersections of the mesh, using PyMesh. 
    These are particularly bad as in this case the area of a face does not correspond to its potentially visible area, as parts of a face can be hidden inside the model, changing the values of $A_z$ and $A_t$.
    As a second step we remove non-surface polygons by computing the visibility of the faces from $1000$ views and drop all non-visible faces of the model using MeshLab. 
    This is primarily done to create cleaner surface meshes by removing unwanted parts of the model, e.g. passenger seats inside planes. 
    This results in a $A_t$ being closer to the actual surface area of the model, while also speeding up the downstream tasks by reducing the number of polygons per model.
    As the first step introduces artifacts in the form of small and irregular meshing, where self-intersections were resolved, we add a third and last step where we regularize the surface meshes by performing an edge-collapse reduction algorithm, again using MeshLab.
    Furthermore, the last step also removes unwanted structures in the meshes, e.g. polygons referring to different textures, which are not relevant for shape information but can influence the viewpoint quality. 
    Fig. \ref{fig:mesh_cleaning} shows details of the model \emph{airplane\_0004} from ModelNet40, which contains self-intersections (top left) and unnecessary polygons (top right).
    The proposed mesh processing resolves self-intersections in the first step and and cleans the meshing in the third step (bottom images).
    We note that this mesh cleaning pipeline does not create perfect watertight surface meshes, but is merely a trade-off between computation time and achieved mesh quality.
    The resulting quality turned out to be adequate for our experiments.
    Providing an algorithm for high quality remeshing is beyond the scope of this paper, and an active field of research on its own.

\section{Distribution Learning}\label{app:distr}
\subsection{Network Architecture}\label{app:distr_arch}
    For predicting the viewpoint quality distribution we use the same feature encoder as for the other tasks, see Section 4.3, and replace the prediction MLPs with 2D decoder networks.
    These decoder networks consists of deconvolution layers to increase the spatial dimensions interweaved with residual blocks to increase the decoder capacity. 
    The whole decoder architecture is as follows:
    \begin{itemize}
        \item 2D deconvolution with filter size $4\times4$ and stride 1,
        \item 2 ResNet blocks with filter size $3\times3$ and depth 2 each,
        \item 2D deconvolution with filter size $4\times4$ and stride 4
        \item 2 ResNet blocks with filter size $3\times3$ and depth 2 each,
        \item 2D deconv layer with filter size $2\times2$ and stride 2,
    \end{itemize}
    with batch normalization and ReLU activation in between all layers.
    The respective spatial dimensions are $(1\times1, 4\times4, 16\times16, 32\times32)$ with feature dimensions $(2048, 1024, 256, 1)$. 
\subsection{Predicted Distributions}\label{app:distr_pred}
\begin{figure}[t]
        \centering
        \includegraphics[width=\linewidth]{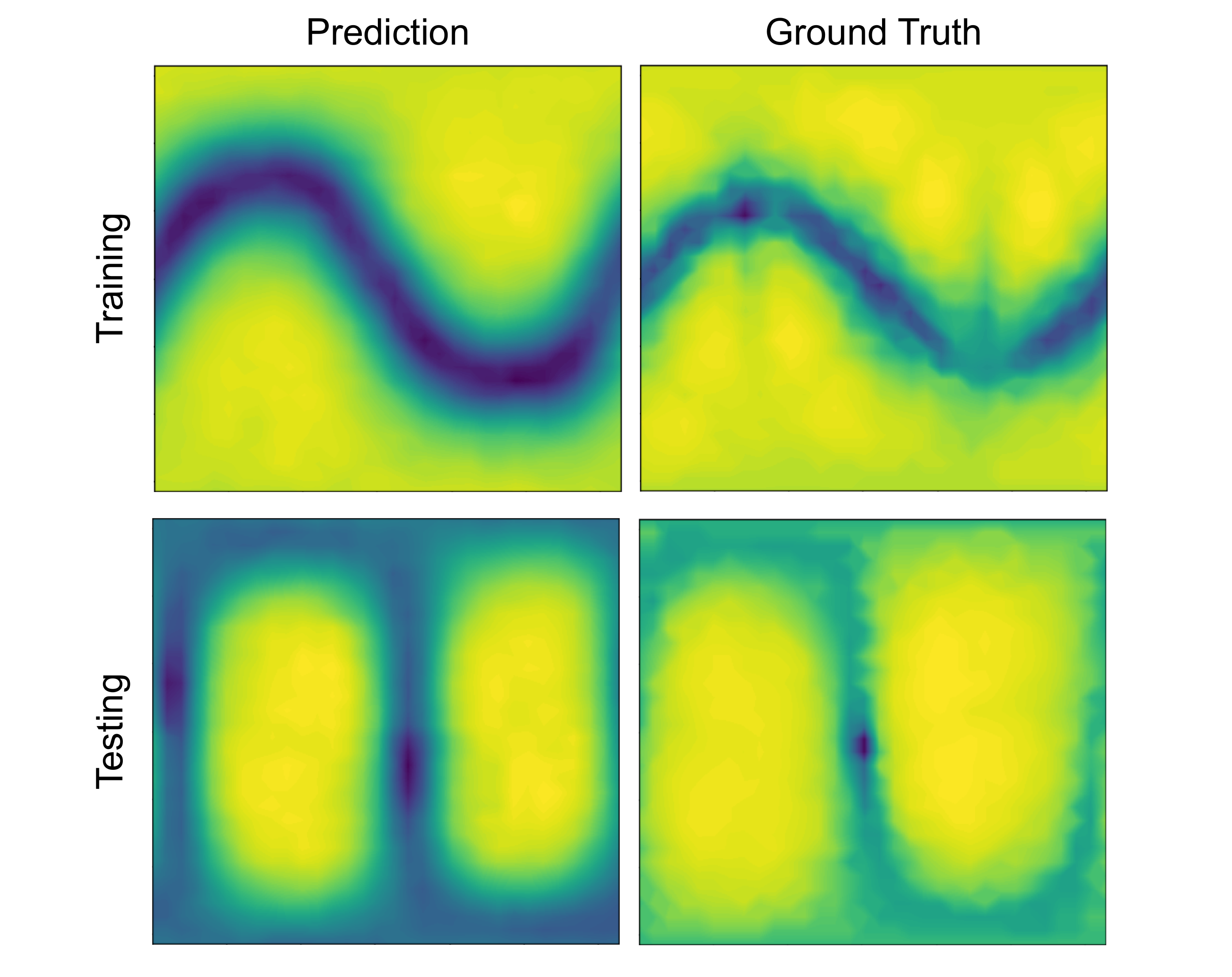}
        \caption{\textbf{Predicted viewpoint quality distributions.}\label{fig:distr_pred}
            Results of the DLDL approach on the category \emph{airplane} for one example from the training set and one example from the test set.}
    \end{figure}

    Examples for predicted viewpoint distributions can be seen in Fig.~\ref{fig:distr_pred}.

     \section{Viewpoint prediction}\label{app:detailed_results}
     Tables \ref{table_vq_results_detail} shows the test results on the different categories for the experiment from Section 5.2. 
     Our combined ML+GL method achieves best performance on almost all categories and viewpoint quality measures.
     The results are comparable on all categories for all viewpoint quality measures, with or without considering statistics, 
     
\section{Dataset}\label{app:data_set}
    We release our training data which contains dense viewpoint quality values for 1k viewpoints on a Fibonacci sphere for \VE, \VR, \VKL and \VMI\ for \textasciitilde 12k models from ModelNet40. 
    For a subset of \textasciitilde 4k models from the categories \emph{airplane, bench, bottle, car, chair, sofa, table and toilet} we additionally provide the cleaned models using our pipeline together with the sampled viewpoint quality values for \VE, \VR, \VKL\ and \VMI. 

\end{document}